\documentclass{emulateapj}
\usepackage{apjfonts}

\newcommand\lsim{\mathrel{\rlap{\lower4pt\hbox{\hskip1pt$\sim$}}
        \raise1pt\hbox{$<$}}}
\newcommand\gsim{\mathrel{\rlap{\lower4pt\hbox{\hskip1pt$\sim$}}
        \raise1pt\hbox{$>$}}}
\newcommand\propsim{\mathrel{\rlap{\lower4pt\hbox{\hskip1pt$\sim$}}
        \raise1pt\hbox{$\propto$}}}

\begin{document}

\title{Prompt Shocks in the Gas Disk Around a Recoiling Supermassive Black Hole Binary}

\author{Zolt\'an Lippai, Zsolt Frei}
\affiliation{Institute of Physics, E\"otv\"os University, P\'azm\'any P. s. 1/A, 1117 Budapest, Hungary}


\author{Zolt\'an Haiman}
\affiliation{Department of Astronomy, Columbia University, 550 West 120th Street, New York, NY 10027, USA}


\begin{abstract}
  Supermassive black hole binaries (BHBs) produced in galaxy mergers
  recoil at the time of their coalescence due to the emission of
  gravitational waves (GWs).  We simulate the response of a thin,
  two--dimensional disk of collisionless particles, initially on
  circular orbits around a $10^6~{\rm M_\odot}$ BHB, to kicks that are
  either parallel or perpendicular to the initial orbital plane.
  Typical kick velocities ($v_{\rm kick}$) can exceed the sound speed
  in a circumbinary gas disk. While the inner disk is strongly bound
  to the recoiling binary, the outer disk is only weakly bound or
  unbound.  This leads to differential motions in the disturbed disk
  that increase with radius and can become supersonic at $\gsim 700$
  Schwarzschild radii for $v_{\rm kick}=500~{\rm km~s^{-1}}$, implying
  that shocks form beyond this radius.  We indeed find that kicks in
  the disk plane lead to immediate strong density enhancements (within
  weeks) in a tightly wound spiral caustic, propagating outward at
  the speed $\sim v_{\rm kick}$. Concentric density enhancements are
  also observed for kicks perpendicular to the disk, but are weaker
  and develop into caustics only after a long delay ($>$one year).
  Unless both BH spins are low or precisely aligned with the orbital
  angular momentum, a significant fraction ($\gsim$ several \%) of
  kicks are sufficiently large and well aligned with the orbital plane
  for strong shocks to be produced. The shocks could result in an
  afterglow whose characteristic photon energy increases with time,
  from the UV ($\sim 10$eV) to the soft X--ray ($\sim 100$eV) range,
  between one month and one year after the merger.  This could help
  identify EM counterparts to GW sources discovered by {\it LISA}.
\end{abstract}

\vspace{\baselineskip}


\vspace{\baselineskip}

\section{Introduction}

The recent break--through in numerical relativity has allowed a direct
computation of the linear momentum flux produced during the
coalescence of a BH binary (Baker et al. 2006, 2007; Campanelli et
al. 2007a,b; Gonz\'alez et al. 2007a,b; Herrmann et al. 2007a,c;
Koppitz et al. 2007).  The resulting final recoil depends on the
masses, orbital parameters, and spins of the BHs, and in special
configurations (i.e. with spins anti-aligned with each other), it can
reach velocities as high as $\approx 4,000~{\rm km~s^{-1}}$
(Gonz\'alez 2007b).  Much of the recent literature focused on the
astrophysical implications of high--velocity kicks, which may displace
or remove supermassive BHs from galactic centers (e.g. Merritt et
al. 2004; Madau \& Quataert 2004; Loeb 2007), and inhibit the growth
of BHs at high redshifts ($z\gsim 6$), where the escape velocities
from low--mass galactic halos is small ($\lsim 100~{\rm km~s^{-1}}$;
Haiman 2004, Yoo \& Miralda-Escude 2004, Shapiro 2005; Volonteri \&
Perna 2005).

Another exciting implication of kicks is that they may help produce an
electromagnetic (EM) counterpart of gravitational wave sources
detected by the future {\it Laser Interferometric Space Antenna} ({\it
LISA}) satellite. The discovery of such a counterpart would constitute
a milestone for fundamental physics and astrophysics (e.g. Kocsis et
al. 2007b). If the BHB is surrounded by a circumbinary gas disk, the
disk will respond promptly (on the local orbital timescale) to such a
kick. If this results in warps or shocks, the disturbed disk could
produce a transient EM signature (Milosavljevi\'c \& Phinney 2005).
The sky localization uncertainty from the {\it LISA} instrument
several weeks prior to merger is typically a few square degrees
(Kocsis et al. 2007a; Lang \& Hughes 2007), and the kick can begin
building up at the end of the inspiral phase, before the final
coalescence (Schnittman et al. 2007).  This will make it possible, in
many cases, to monitor a few square--degree area on the sky prior to,
during, as well as immediately following the coalescence of BHs in the
mass range $\sim (10^5$--$10^7)\,{\rm M_\odot}/(1+z)$ at redshifts out
to $z\sim 3$, and search for a prompt transient signature associated
with the kick (Kocsis et al. 2007b).

In the context of producing a prompt EM counterpart, we expect that
the {\it direction}, in addition to the magnitude of the kick, will be
important. Naively, one expects that a kick within the plane of any
circumbinary disk will be more likely to cause density enhancements
and ``light up'' the disk than a kick perpendicular to it.  The kick
direction can be clearly important on larger scales, as well, and
affect phenomena that would occur on long timescales ($\gg$ years)
after the merger. For example, the angle relative to a large--scale
galactic disk can determine whether a kicked BHB ends up outside the
galaxy or not, and perhaps also whether shocks are produced when a BHB
plunges into such a large--scale disk.  While these issues merit
investigation, in this {\em Letter} we focus on the smaller scales and
shorter timescales ($\lsim$ months) that are relevant to prompt {\it
LISA} counterparts.  

Schnittman \& Buonanno (2007; hereafter SB07) used the ``effective
one--body approach'' and derived a scaling formula that yields the
recoil velocity vector for arbitrary mass ratios and spin
vectors. While their results have not been tested for generic spins,
they agree well (to within 20-30\%) with numerical results in those
special configurations where they were tested (including numerical
calculations for configurations with the spins parallel or at
intermediate angles with respect to the orbital plane; see Campanelli
et al. 2007a,b, Gonz\'alez et al. 2007a; Herrmann et al. 2007b, Tichy
\& Marronetti 2007).

In this {\em Letter}, we investigate the response of a circumbinary
disk to the kick, by following the perturbed, Keplerian orbits of
collisionless massless test particles around the recoiling BHB.  The
purpose of this exercise is to demonstrate that prompt shocks or
strong density enhancements are likely to arise when the kick is
aligned with the plane of the circumbinary disk, whereas they may be
less likely for highly inclined kicks.  We will then use the formula
of SB07 for the dependence of the kick speed and direction on the mass
ratio and spins of the BHs, to argue that a significant fraction
($\gsim$ few \%) of kicks may be both sufficiently large and well
aligned with the orbital plane for strong shocks to be produced within
a few weeks after coalescence.

\section{Circumbinary Disks}

We begin with the assumption that the kicked BHB is surrounded by a
rotationally supported, geometrically thin gaseous circumbinary disk.
The idealized case of a smooth axisymmetric accretion disk, aligned
with the binary's orbital plane (Bardeen \& Peterson 1975), with a
simple vertical structure can be described by its density and
temperature profiles ($\rho(r)$, $T(r)$), as well as the scale height
$H(r)$.  This minimal information is needed (i) to compute the angle
extended by the disk as viewed from the center, and (ii) to examine
whether the kick is supersonic ($v_{\rm kick} > c_s$; a necessary
condition for shocks to occur). In addition, we need to know the
overall disk size (necessary to assess what fraction of the disk gas
remains bound to the recoiling BHB), and whether the disk mass is
comparable to the BHB mass (necessary to assess whether the BHB is
slowed down by the disk).

Unfortunately, fully self-consistent and stable accretion disk models
around supermassive BHs are both difficult to produce, and require
many ad--hoc assumptions (see, e.g., the review by Blaes 2007 for a
list).  While variants of the parametric so--called $\alpha$--disk
models (Shakura \& Sunyaev 1973) have been successfully calibrated
against observations of stellar--mass objects, observations of active
galactic nuclei (AGN) have not led to similarly robust constraints on
disk models around supermassive BHs.  Nevertheless, several authors
have discussed the possible behavior of a circumbinary $\alpha$--disk
\citep[e.g.][]{an02,mp05,dot06}.  A key quantity is the critical
orbital semi--major axis $a_{\rm crit}$ at which the time--scale for
the decay of the orbit due to gravitational radiation becomes shorter
than the time--scale for viscous evolution (Begelman et al. 1980).
Once the binary separation shrinks below this value at $t_{\rm crit}$,
the gas outside will not evolve significantly before the BHs coalesce.
Most of the ``kick'' momentum is accumulated during the final stages
of coalescence, on a rapid time--scale \citep{sb07}, hence a snapshot
of the disk at $t_{\rm crit}$ outside $\sim a_{\rm crit}$ should still
describe the disk around the kicked binary.  Prior to $t_{\rm crit}$,
the torques from the binary may create a central cavity nearly devoid
of gas within the radius $r_{\rm cavity}\sim 2a_{\rm crit}$ \citep[for
a nearly equal--mass binary, e.g.][]{al94,mm06}, or a narrower gap
around the orbit of the lower--mass BH in the case of unequal masses
$q \equiv M_1/M_2 \ll 1$ \citep[e.g.][]{an02}.  In the latter case,
the lower--mass hole will ``usher'' the gas inward as its orbit
decays, so by the time of coalescence, this gas may accrete onto the
more massive BH, again leaving a nearly empty central cavity of radius
$a_{\rm crit}$.

In the inner regions of a disk around a supermassive BH, the dominant
vertical support is expected to be radiation pressure (as opposed to
gas pressure), and electron scattering dominates the opacity over
absorption \citep[e.g.][p. 440]{st83}.  \citet{mp05} presented an
explicit model for a circumbinary disk with a central cavity, for the
case of a nearly equal-mass binary at $t_{\rm crit}$. They find, under
the above conditions for the inner disk ($r_{\rm cavity} = 2a_{\rm
crit}$),
\begin{eqnarray}
\nonumber
r_{\rm cavity} = 117 r_{\rm S} 
\alpha_{-1}^{-0.34}
(\eta_{-1}/\dot{m})^{0.24}
M_6^{0.08}
[4q/(1+q)^2]^{0.42},
\end{eqnarray}
where $M_6$ denotes the total BHB mass $M_1+M_2$ in units of $10^6{\rm
M_\odot}$, $r_{\rm S}=2GM/c^2$ is the Schwarzschild radius,
$0.1\alpha_{-1}$ is the effective $\alpha$--parameter relating the
kinematic viscosity to the gas (not total) pressure, $\dot{m}$ is the
mass accretion rate at $r_{\rm cavity}$ in units of the Eddington rate
$\dot{M}_{\rm Edd}$ (i.e. the accretion rate that would produce the
Eddington luminosity with the radiative efficiency $0.1\eta_{-1}$),
and $q=M_1/M_2\leq 1$ is the mass ratio (note that we neglect here the
dependence on additional dimensionless coefficients of order unity).

The thickness of the disk at its inner edge $r_{\rm cavity}$ is given
by
\begin{equation}
\frac{h}{r} = 0.46 
M_6^{-0.12}
[4q/(1+q)^2]^{-1.84}
\alpha_{-1}^{0.76}
(\dot{m}/\eta_{-1})^{2.43}
\end{equation}
and the mid--plane gas temperature is
\begin{equation}
T = 1.7\times 10^6 
M_6^{-0.28}
[4q/(1+q)^2]^{-0.49}
\alpha_{-1}^{0.19}
(\dot{m}/\eta_{-1})^{0.86}\, {\rm K}.
\end{equation}

The ratio of radiation and gas pressure is given by
\begin{equation}
\beta \equiv \frac{P_{\rm rad}}{P_{\rm gas}} = 2600
M_6^{-0.04}
[4q/(1+q)^2]^{-1.84}
\alpha_{-1}^{1.67}
(\dot{m}/\eta_{-1})^{4.25}.
\end{equation}

Note that the disk is marginally thin at $r_{\rm cavity}$.  The
expected radial profile of such a disk outside $r_{\rm cavity}$ was
discussed by \citet{gt04}.  In the stable case considered by
\citet{mp05}, when the viscosity is proportional to the gas (rather
than the total) pressure, the mid--plane temperature and density, and
surface density vary with radius as $T\propto r^{-9/10}$, $\rho\propto
r^{-3/5}$, and $\Sigma\propto r^{-3/5}$.  The ratio of gas/total
pressure varies as $\beta\equiv P_{\rm gas}/P_{\rm tot}\propto
r^{21/10}$ (in the limit of $\beta\ll 1$).  Inside the transition
radius where $\beta\approx 1$,
\begin{equation}
r_{\rm tr}\approx 1.9\times 10^3 
M_6^{2/21}
\alpha_{\rm tot,-1}^{2/21} 
(\dot{m}/\eta_{-1})^{16/21}
r_{\rm S},
\end{equation}
the scale height $h$ is constant, or $h/r\propto 1/r$ (here
$\alpha_{\rm tot,-1}$ relates the kinematic viscosity to the total
pressure). This further assumes that the disk is optically thick.
Beyond $r_{\rm tr}$, gas pressure becomes dominant for vertical
support, and the scale--height begins to rise as $h\propto r^{21/20}$.
Somewhat farther out, beyond $\sim 10,000 r_{\rm S}$ (for $M_6\sim
1$), the disk will become self--gravitating and subject to
Toomre--instability.  There are both theoretical and observational
reasons to suspect that disks in real AGN extend to larger radii
(Goodman 2003; Blaes 2007). However, note that the orbital time--scale
at this radius is $\approx$2 years, so that in the context of prompt
{\it LISA} counterparts, we are not interested in radii outside this
annulus.

\begin{figure}[tbh]
\centering
\mbox{\includegraphics[width=7.5cm]{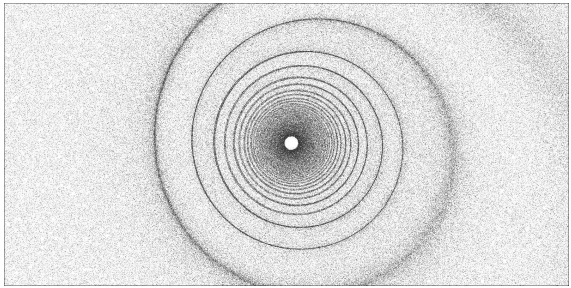}}
\caption{\label{fig:spirals} The top view of a disk around an
  $M_1+M_2=10^6~{\rm M_\odot}$ BH binary, which recoiled within the
  disk plane at the velocity of $v_{\rm kick}=500~{\rm km~s^{-1}}$,
  oriented vertically upward in the diagram.  The disk is initially
  assumed to have an inner edge at $r_{\rm in}=100 r_{\rm S}$, and is
  shown here out to a radius of $r_{\rm in}=5,000 r_{\rm S}$, at a
  time $t=90$ days after the kick.  A similar tightly wound spiral
  caustic patterns develops already in less than a month.  The
  dark/light shades correspond to regions of high/low density, the
  low--density regions being similar to the initial surface density,
  and the high--density regions about 10 times overdense.}
\end{figure}

To summarize, based on the above, we adopt the following simplified
picture for the disk around a fiducial $M_1+M_2=10^6~{\rm M_\odot}$
binary.  The disk has an inner edge at $100 r_{\rm S}$ (inside which
it is empty) and an outer edge at $10,000 r_{\rm S}$ (outside which
there may still be gas, but it evolves slowly and we do not follow
it).  The scale--height and temperature at the inner edge is
$h/r=0.46$ and $T=1.7\times10^6$K, respectively.  The scale--height
remains constant with radius out to $2,000 r_{\rm S}$, beyond which it
increases nearly linearly ($h\propto r^{21/20}$). The temperature
varies with radius as $T\propto r^{-9/10}$.

The important features of such a disk (as well as other proposed
variants of $\alpha$-disks) are the following: (i) orbital motions in
thin disks are supersonic (Pringle 1981), so that the gas is
susceptible to shocks if disturbed; (ii) at the relevant radii outside
$100r_{\rm S}$ the viscous time--scale is long, so that the orbits are
near Keplerian; (iii) gas near the inner edge of disk is tightly bound
to the kicked BHB ($v_{\rm orbit}\sim 3\times 10^4$ km/s), but the
outer edge ($v_{\rm orbit}\sim 3\times 10^3$ km/s) can be marginally
bound, or even unbound, for large kicks (the approximate condition for
being bound is $v_{\rm orbit}\gsim 2.4 v_{\rm kick}$), and (iv) the
total disk mass within $10,000 r_{\rm S}$ is much less than the BHB
mass, which justifies ignoring the inertia of the gas bound to the
BHB.

\section{The Response of the Disk to Kicks}

For a quantitative assessment of the disk's response to the kick, we
employ the following approximation: the disk particles are assumed to
be massless, collisionless, and initially on co-planar, circular
orbits. The kick simply adds the velocity $\vec{v}_{\rm kick}$ to the
instantaneous orbital velocity of each particle (in the inertial frame
centered on the BHB).  We used $N=10^6$ particles, distributed
randomly and uniformly along the two--dimensional surface of the disk.
The kick velocity was varied between $500~{\rm km~s^{-1}} < v_{\rm
kick} < 4,000~{\rm km~s^{-1}}$, and directed either perpendicular or
parallel to the initial disk plane.

Figure~\ref{fig:spirals} shows, as an example, a face--on view of the
surface density of the disk $90$ days after a kick with $v_{\rm kick}
= 500~{\rm km~s^{-1}}$ in the plane of the disk ($i_{\rm
kick}=0^\circ$).  The sharp, tightly wound spiral features clearly
seen in the figure trace the locus of points where particles cross
each other, corresponding formally to a density caustic.  The spiral
caustic first forms at $\sim 30$days, and then propagates outward at a
speed of $\approx 500~{\rm km~s^{-1}}$.  This behavior can be roughly
understood as follows: at a given radius $r_c$, the caustic forms at
the time $t_{\rm c}$ when the radial epicyclic motions from two
neighboring annuli, separated by the epicyclic amplitude $\sim (v_{\rm
kick}/v_{\rm orbit}) r_c$, overlap (this predicts $r_c\sim t_{\rm c}
v_{\rm kick}$).  We also found that as $v_{\rm kick}$ is increased,
the spiral caustics spread farther out.  Eventually, for kicks strong
enough so that a significant fraction of the disk is unbound, the
caustics loose their coherent spiral patterns, and develop complex
two--dimensional shapes.

\begin{figure}[tbh]
\centering
\mbox{\includegraphics[width=7.5cm]{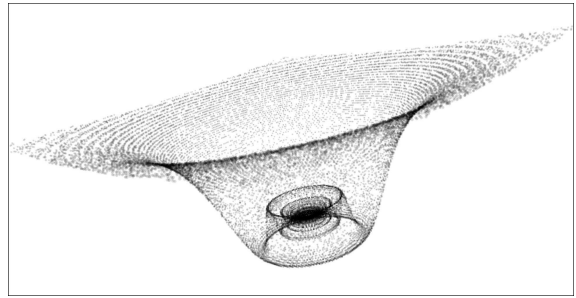}}
\caption{\label{fig:sombrero} The aerial view of a disk as in
Figure~\ref{fig:spirals}, except here the kick is oriented
perpendicularly to the disk, and the snapshot is taken at $t=1$
week. For visual clarity, the graphic contrast was increased relative
to Fig.~\ref{fig:spirals}, and the diagram stretched by a factor of 10
along the axis perpendicular to the disk plane.  The density
distribution is azimuthally symmetric, and while there are mild
concentric density fluctuations, strong enhancements (i.e. caustics)
develop only after a delay of $\approx$ one year.}
\end{figure}

Figure~\ref{fig:sombrero} shows a side view of the 3D particle density
one week after a kick with the velocity $v_{\rm kick} = 500~{\rm
km~s^{-1}}$ perpendicular to the disk ($i_{\rm kick}=90^\circ$).  The
density profile in this case remains azimuthally symmetric, but still
develops concentric rings of density fluctuations.  The important
difference from the parallel kick case is that the density
enhancements are much weaker (at the ten percent level).  By examining
the time--evolving radial cross--sections, we have verified that sharp
density enhancements, i.e. true caustics caused by the orbit--crossing of
particles, first appear only after one year, and involve a smaller
fraction of the disk particles.

More generally, kicks can be expected to occur in a direction
intermediate between the two extremes shown in Figs.~
\ref{fig:spirals} and \ref{fig:sombrero}.  Following a kick at an
arbitrary inclination angle, each particle still remains on a Kepler
orbit. Each new orbit will follow an ellipse in a plane that is tilted
by an angle of order between $\pm v_{\rm kick}/v_{\rm orbit}$ about
the axis connecting the instantaneous position of the particle with
the central BHB.  Hence, after an orbital time, the particles of the
disk will be smeared vertically, effectively thickening the original
disk.  The two--dimensional simulations we performed do not conclusive
tell us a critical kick inclination angle for prompt density caustics
to be produced -- this would require three--dimensional simulations,
following the orbits of particles in a 3D disk of finite
thickness. However, since the prompt, strong density enhancements
appear to develop in the plane of the kick, near the inner edge of the
disk, one may conjecture that this critical inclination angle is given
roughly by $i_{\rm kick}\lsim \arctan(h/r)$ (with $h/r=0.46$ evaluated
at $r_{\rm cavity}$).

\begin{figure}[tbh]
\centering
\mbox{\includegraphics[height=4.2cm]{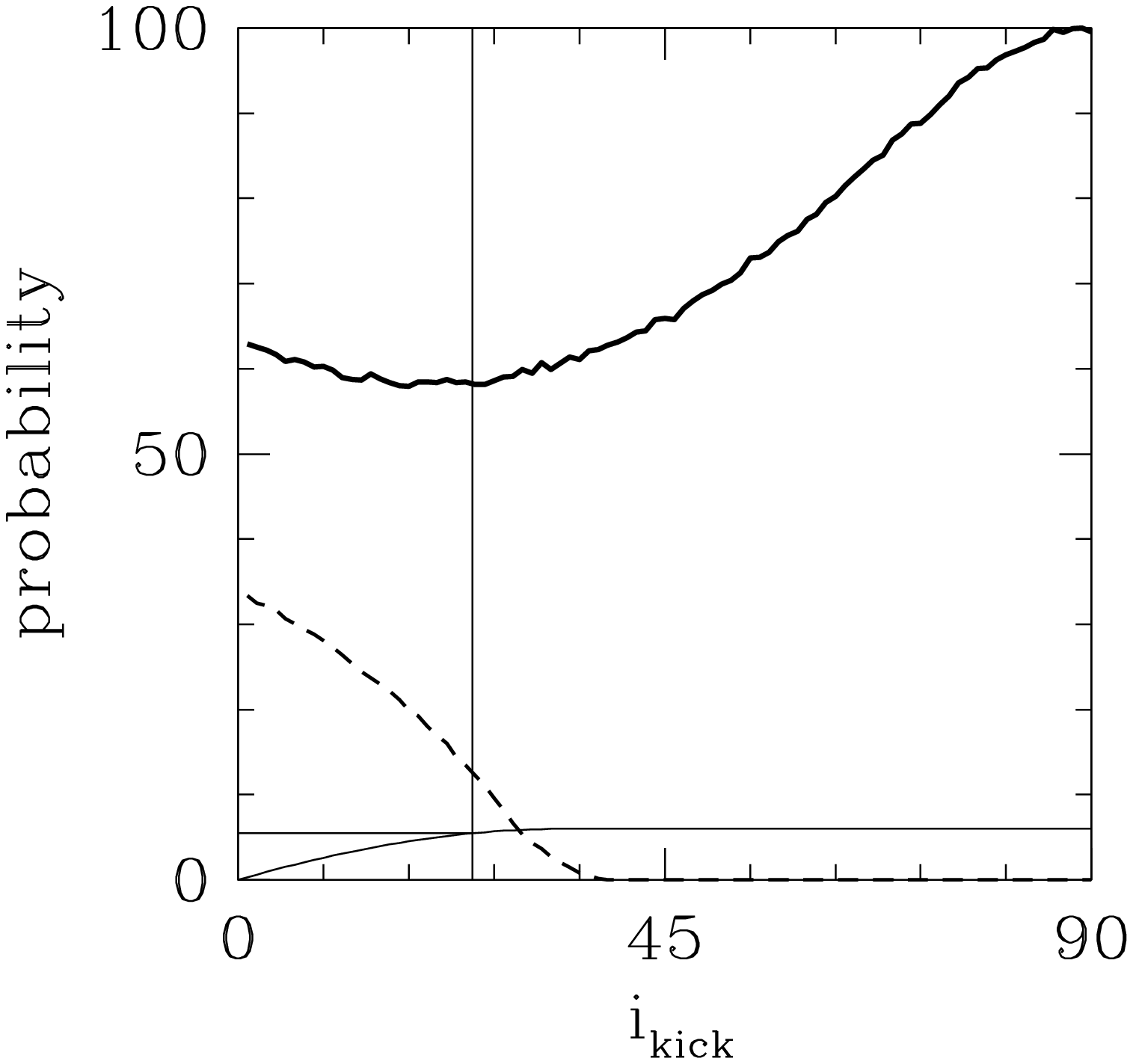}\includegraphics[height=4.2cm]{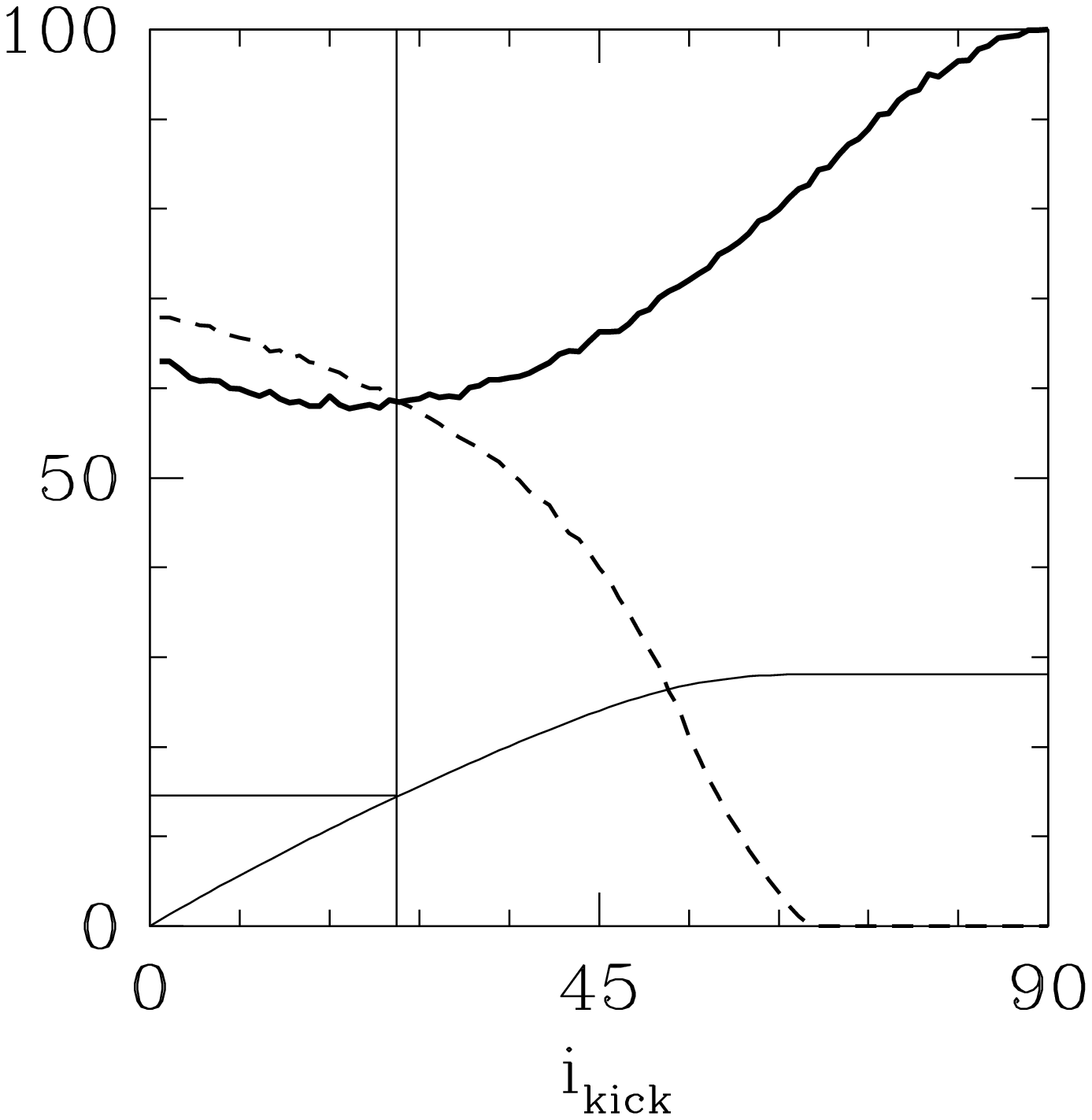}}
\caption{\label{fig:angles} The thick solid curves in both panels show
  the (arbitrarily normalized) differential probability distribution
  of kick inclinations relative to the disk plane, using the SB07
  formula to compute $\vec{v}_{\rm kick}$. Both panels assume an
  equal--mass binary ($q=1$; we find results changed only by a
  few percent for $q=0.1$).  We assume random spin orientations.  On
  the left panel we assume dimensionless spins of $a_1=a_2=1$, while
  on the right panel $a_1=a_2=0.9$.  The dashed curves show the
  fraction of kicks at a given inclination angle that have a component
  in the disk exceeding a fixed threshold ($v_{\rm kick,\parallel}\geq
  500~{\rm km~s^{-1}}$ in the left panel, and $300~{\rm km~s^{-1}}$ in
  the right panel).  The thin curves indicate the cumulative
  probability of sufficiently fast kicks with inclinations below a
  given value; the horizontal line shows the fraction of such kicks
  with inclinations $\leq 24^\circ$ (6 and 15\% percent in the left
  and right panels). }
\end{figure}

In Figure~\ref{fig:angles}, we show the distribution of kick angles,
taken from SB07, assuming that both spins are randomly oriented.  We
require further that the component of $v_{\rm kick}$ within the
orbital plane exceed $3-500~{\rm km~s^{-1}}$ ($\approx 3-5\times$ the
sound speed; see justification in \S~\ref{sec:discuss} below).  A
strong kick within the orbital plane is produced if the spins are
large and parallel to the angular momentum, but anti--aligned with
each other.

The main conclusion to draw from Figure~\ref{fig:angles} is that
caustic--producing kicks are not rare -- in the cases shown in the
panels, they occur at least a few percent of the cases.  This
conclusion remains true unless (i) the spins of the BHs is
significantly below the maximal Kerr value ($a\lsim 0.9$), or (ii) the
spins are significantly aligned with each other.  The latter may be
expected if the spin of both BHs results from accretion from the same
disk \citep{brm07}.

The criteria defined above for producing caustics may turn out very
conservative.  For example, prompt caustics could develop when
particles from different layers of the disk cross along their orbits.
A plausible weaker criterion for shocks is that the kick--induced
tilts of the orbits do not thicken the disk beyond its original
scale--height.  Apart from trigonometric factors of order unity that
describe its azimuthal dependence, the tilt angle is given by $i_{\rm
tilt} \sim (v_{\rm kick}/v_{\rm orbit}) i_{\rm kick}$; requiring only
$i_{\rm tilt} \lsim \arctan(h/r)$, or $i_{\rm kick} \lsim (v_{\rm
orbit}/v_{\rm kick})\arctan(h/r)$, density caustics could arise even
for nearly perpendicular kicks.  In follow--up work, we will employ
three--dimensional simulations, to clarify the relevant caustic
formation criterion.

\section{Discussion}
\label{sec:discuss}

The main result of this {\it Letter} is that strong density
enhancements can form promptly after a supersonic kick in the
plane of the circumbinary disk, within a few weeks of the coalescence
of a $\sim 10^6~{\rm M_\odot}$ BHB.  Because the disk is cold, and 
caustics (Fig.~\ref{fig:spirals}) are formed when particles first
cross each other along their orbits, this implies that corresponding
shocks could occur in a gas disk.  For hydrodynamical shocks to occur
within a finite--pressure gas, the relative motions $v_{\rm c}$
between the neighboring particles that produce the caustic must exceed
the sound speed.  At the outermost radius where the disk is marginally
bound to the BHB, one expects $v_{\rm c}\sim v_{\rm kick} \sim v_{\rm
orbit}$; relative motions will be slower further inside.  The relative
speed should roughly correspond to covering the epicyclic amplitude
$\sim (v_{\rm kick}/v_{\rm orbit}) r_c$ in the caustic--formation time
$t_{\rm c}\sim r_c/v_{\rm kick}$, yielding $v_c\sim v_{\rm
kick}^2/v_{\rm orbit}$.  For $v_{\rm kick}= 500 {\rm km~s^{-1}}$, this
predicts $v_{\rm c} \sim 25 {\rm km~s^{-1}} (r/1000r_{\rm S})^{1/2}$;
we have verified in our simulations that particles cross the caustics
with speeds at about $\sim 30\%$ above this predicted value.  Compared
with the sound speed $c_s\approx 25 {\rm km~s^{-1}} (r/1000r_{\rm
S})^{-9/20}$, this suggests that the density waves produced by the
kick in the gas beyond $\sim 700r_{\rm S}$ will indeed steepen into
shocks.  We also found that the inclination of the kick may be
important in determining the strength and timing of such shocks --
perpendicular kicks would only produce weaker density enhancements, at
least until a delay of about a year.  A non--negligible fraction
($\gsim$ several \%) of kicks could, however, be sufficiently large
and well aligned with the orbital plane for shocks to be produced
within a few weeks after coalescence.

The nature of the emission resulting from the shocks or density
enhancements will have to be addressed in future work, by computing
the heating rate at the spiral shocks, and modeling the overall disk
structure and vertical radiation transport.  However, the disk's
luminosity resulting from the kick, $L_{\rm kick}$ could be a
non--negligible fraction of the BHB's Eddington luminosity, and
therefore potentially observable (Kocsis et al. 2007b).  If $M_{\rm
shock}=f_{\rm shock}(M_1+M_2)$ is the mass of the shocked gas that is
heated to temperatures corresponding to $v_{\rm shock}$, and $t_{\rm
shock}$ is the time--scale on which the corresponding thermal energy
is converted to photons, then $L_{\rm kick} \approx (1/2) M_{\rm
shock} v_{\rm shock}^2 / t_{\rm shock}$, and $L_{\rm kick}/L_{\rm Edd}
\approx 0.1 (f_{\rm shock}/10^{-5}) (v_{\rm shock}/500~{\rm
km~s^{-1}})^2 (t_{\rm shock}/{\rm 1 month})^{-1}$.  We may also
speculate on the spectral evolution of the ``kick
after--glow''. Assuming $M_{\rm shock} \propto \Sigma r dr \propto r
^{19/10}$ (with $\Sigma \propto r^{-3/5}$ and $dr\propto r^{1/2}$, the
epicyclic amplitude), and $v_{\rm shock} \approx v_c$, we find $L_{\rm
kick} \propto r^{24/10}$. This suggests that the luminosity may be
dominated by the outermost shocked shells. The spectrum will then peak
at the characteristic photon energy corresponding to $k T_{\rm
shock}\propto v_{\rm c}^2 \propto v_{\rm orbit}^{-2} \propto r$.  The
shocks could therefore result in an afterglow, starting from $r_{\rm
cavity}/v_{\rm kick}\sim 30$ days, first peaking in the UV band ($\sim
10$eV), and then hardening to the soft X--ray ($\sim 100$eV) range
after one year. The detection of such an afterglow would help identify
EM counterparts to GW sources discovered by {\it LISA}.

Our results need to be verified in three--dimensional simulations that
resolve the orbits within a finite disk thickness.  A realistic disk
model, incorporating gas dynamics, is needed to study the
correspondence between collisionless caustics and gaseous
shocks. Finally, the SB07 formula needs to be confirmed for generic
spin configurations.  Nevertheless, our results do suggest that kicks
due to gravitational waves may produce a prompt EM signal.


We thank A. Buonanno, M. Milosavljevi\'c, K. Menou, and B. Kocsis for
useful discussions, and M. Milosavljevi\'c for suggesting the
arguments based on epicycles. This work was supported by the Pol\'anyi
Program of the Hungarian National Office for Research and Technology
(NKTH) and by NASA grant NNG04GI88G (to ZH).

\end{document}